\documentclass[final,4p,times]{elsarticle}

\usepackage{amssymb}

\usepackage[colorlinks,linkcolor=red,anchorcolor=blue,citecolor=green]{hyperref}
\usepackage{amsthm}
\usepackage{amsmath}
\usepackage{subfigure}
\usepackage{anysize}
\marginsize{2.2cm}{2.2cm}{0cm}{1.5cm}
\selectfont

\begin{document}
\newtheorem{theorem}{\indent Theorem}[section]
\newtheorem{proposition}[theorem]{\indent Proposition}
\newtheorem{definition}[theorem]{\indent Definition}
\newtheorem{lemma}[theorem]{\indent Lemma}
\newtheorem{remark}[theorem]{\indent Remark}
\newtheorem{corollary}[theorem]{\indent Corollary}
\newcommand{\noin}{\noindent}
\newcommand{\ba}{\begin{array}}
\newcommand{\ea}{\end{array}}
\newcommand{\fr}{\frac}
\newcommand{\beq}{\begin{equation}}
\newcommand{\eeq}{\end{equation}}
\newcommand{\ben}{\begin{enumerate}}
\newcommand{\een}{\end{enumerate}}
\newcommand{\bit}{\begin{itemize}}
\newcommand{\eit}{\end{itemize}}
\newcommand{\bmath}[1]{\mbox{\boldmath$ #1 $}}
\newcommand{\ds}{\displaystyle}
\newcommand{\figcaption}
{\renewcommand{\thesection}{Figure captions}}
\newcommand{\namelistlabel}[1]{\mbox{#1}\hfil}
\newenvironment{namelist}[1]{\begin{list}{}
{\let\makelabel\namelistlabel
\settowidth{\labelwidth}{#1}
\setlength{\leftmargin}{1.1\labelwidth}}
 }{\end{list}}
\newtheorem{df}{Definition}
\newtheorem{lem}{Lemma}
\newtheorem{prop}{Proposition}
\theoremstyle{remark}
\begin{frontmatter}



\title{Integrable discrete massive Thirring model}

 \author[label1]{Junchao Chen}
\author[label2]{Bao-Feng Feng \corref{cor1}}\ead{baofeng.feng@utrgv.edu}
 \address[label1]{Department of Mathematics,
Lishui University, China}
\address[label2]{School of Mathematical and Statistical Sciences,
The University of Texas Rio Grade Valley, USA}

\begin{abstract}
In this paper, we are concerned with integrable semi- and fully discrete analogues of the massive Thirring model in light core coordinates. By using the Hirota's bilinear approach and the KP reduction method, we propose both the semi- and fully discrete massive Thirring models and construct their multi-bright soliton solutions.
\end{abstract}


\end{frontmatter}

\section{Introduction}
The massive Thirring (MT) model
\begin{eqnarray}
&&  \displaystyle \mathrm{i} u_x + v +  u|v|^2 =0\,,  \label{MTa} \\ [5pt]
&& \displaystyle \mathrm{i} v_t + u +  v|u|^2  =0 \label{MTb} \,,
\end{eqnarray}
 was derived by Thirring in 1958  \cite{Thirring} in the context of
general relativity. It represents a relativistically invariant nonlinear Dirac equation in the space of
one dimension. It is one of the most beautiful and remarkable solvable field theory models. Its complete integrability was firstly approved by
Mikhailov \cite{Mikhailov} and Orfanidis \cite{Orfanidis} independtly.  The inverse scattering transform for the MT model was studied by
Kuznetsov and Mikhailov \cite{Kuznetsov} and many others \cite{Kawata79,WadatiMT83,KaupLakoba,Villarroel:1991,DmkitryMTIST}. The Darboux transformation, B\"acklund transformation of the MT model and its connection with other integrable systems have been extensively investigated  \cite{KaupNewell,Lee:1993,Lee:1994,NijhoffMT1,NijhoffMT2,Prikarpatskii:1979,Prikarpatskii:1981,Franca13,DegasperisMT}.

Since the pioneer work by Date \cite{Date}, the soliton solutions to the MT model including the ones with nonvanishing background were constructed by many authors
\cite{Shnider84,Alonso:1984,BarashenkovGetmanov:1987,BarashenkovGetmanovKovtun:1993,BarashenkovGetmanov:1993,Talalov:1987,Vaklev:1996}.

Regarding the rogue wave solutions of the MT model, the first-order rogue wave has been constructed through the Darboux method with a matrix version of the Lagrange interpolation method \cite{Degasperis2015Bragg,Degasperis2015darboux}.
Applying this fundamental solution to the coupled mode equations has shown that combining electromagnetically induced transparency with Bragg scattering four-wave mixing may yield rogue waves at low powers.
The higher-order rogue wave solutions have been investigated by using the n-fold Darboux transformation, and the explicit formulas up to third-order with their patterns have been provided in detail \cite{guo2017rogueMT}.
By virtue of the nonrecursive Darboux transformation method, general super rogue wave solutions have been obtained and their structure analyses reveal that rogue waves properties of both components in the coupled MT model is same as that in scalar nonlinear systems except for the spatiotemporal distributions \cite{ye2021super}.

In \cite{CFBilinearMT}, the connection between the MT model to the KP-Toda hierarchy was clarified and various soliton soliton solutions such as bright, dark and breather solutions were constructed. Based on this finding, the first- and second-order algebraic solutions to the MT modePelinovskyMTPREl. were derived in \cite{PelinovskyMTPRE}. In \cite{CYFRogueMT}, general rogue wave solutions in terms of determinants whose matrix elements are elementary Schur polynomials in the MT model were derived by using the KP reduction method. It was also shown that the patterns of rogue waves are associated with the root structures of the Yablonskii–Vorob'ev polynomial hierarchy through a linear transformation.
Recently, the rogue wave solutions to the MT model were investigated in \cite{HeMT2015,ShihuaMT} by the Darboux transformation.

The study of discrete integrable systems has been attracted much attention in the past as well. As a result, integrable discrete analogues of many soliton equations such as the KdV equation has been proposed \cite{Disbook}. However, unlike other soliton equations, the study of integrable discretizations of the MT model is relatively rare.  Nijhoff {\it et al.} gave the integrable discretization of the MT model in light cone coordinates \cite{NijhoffMT1,NijhoffMT2}. Until recently, Pelinovsky {\it et al.} proposed a semi-discrete integrable MT model
in laboratory coordinates \cite{DmkitryMTdiscrete} and studied its solution solution via Darboux transformation \cite{DmitryXu2019}.
Most recently, we derive a set of bilinear equations corresponding to either the vanishing or non-vanishing boundary conditions and further construct general bright and dark soliton solutions via the KP reduction method \cite{CFBilinearMT}.  We found that the bilinear equations, as well as multi-bright soliton solution, of the MT model can be reduced from the two-component KP-Toda hierarchy.

The purpose of the present paper is to construct semi- and fully discrete MT model via the combination of Hirota's bilinear method \cite{Hirotabook} and the KP reduction method \cite{JM}.
By using this method, we have constructed general soliton solutions to many soliton equations such as the vector Nonlinear Schr\"odiner equation \cite{FengvNLS} and the complex short pulse equation
\cite{FengShen_ComplexSPE,FMO_ComplexSPE}. In the present paper, we propose the integrable semi- and fully discrete MT model in light cone coordinates, which is different from
the ones in \cite{NijhoffMT1,NijhoffMT2} and has no square singularity. The remainder of the paper is organized as follows. We state the main results: the discrete MT model and the corresponding multi-soliton solutions in section 2.  Then,  section 3 are devoted to the bilinear equations and the reduction procedure from the KP-Toda hierarchy for the semi- and fully discrete MT model, respectively. The paper is concluded in section 4 by some comments and further topics.

\section{Main results}

In this section, we list the main results for the semi-discrete and fully discrete MT model, along with their multi-soliton solutions by two theorems.

\begin{theorem}
We propose a semi-discrete analogue of the MT model (\ref{MTa})-- (\ref{MTb})
\begin{eqnarray}
&&  \displaystyle \frac{{\rm i}}{a}   (u_{k+1} - u_{k}) + v_{k+1} +   u_{k+1} v_{k+1} \tilde{v}_{k} =0\,,  \label{semi-MTa} \\ [5pt]
&& \displaystyle \mathrm{i} v_{k,t} + u_k +  v_k u_k \tilde{u}_k  =0 \label{semi-MTb} \,,
\end{eqnarray}
where $a$ is the lattice spacing parameter along $x$-direction and $k$ is the discrete lattice variable such that $u_k=u(x=ka,t)$, $v_k=v(x=ka,t)$.
The semi-discrete MT model (\ref{semi-MTa})-- (\ref{semi-MTb}) admits the following multi-soliton solution
\begin{equation} \label{bt_tran1}
u_k=\frac{g_k}{\tilde{f}_k}\,, \quad
v_k=\frac{h_k}{f_k}\,, \quad
\tilde{u}_k=\frac{\tilde{g}_k}{f_k}\,, \quad
\tilde{v}_k=\frac{\tilde{h}_k}{\tilde{f}_k}\,,
\end{equation}
where
\begin{equation}
f_k=f^0_{1,k},\ \ \tilde{f}_k=f^0_{0,k},\ \ g_k= -{\rm i} g^1_{0,k} ,\ \  h_k=g^0_{0,k}\,\ \ \tilde{g}_k={\rm i}\bar{g}^{-1}_{0,k},\ \ \tilde{h}_k=-\bar{g}^0_{0,k} .
\end{equation}
Here $f^m_{n,k}$, $g^m_{n,k}$ and $\tilde{g}^m_{n,k}$ are $2N \times 2N$ and $(2N+1) \times (2N+1)$ determinants defined
by
\begin{equation}
f^{m}_{n,k} =\left\vert
\begin{array}{cc}
A^{k}_n & I \\
-I & B_m\end{array}\right\vert \,, \quad
g^{m}_{n,k} =
\left\vert
\begin{array}{ccc}
A^{k}_n & I & (\Phi_n^{k})^T \\
-I & B_m & \mathbf{0}^{T} \\
\mathbf{0} & -\bar{C}_m & 0\end{array}\right\vert  \,, \ \ \
\tilde{g}^{m}_{n,k} =\left\vert
\begin{array}{ccc}
A^{k}_{n-1} & I & \mathbf{0}^{T} \\
-I & B_m & (C_m)^{T} \\
-\bar{\Phi}^{k}_n & \mathbf{0} & 0\end{array}\right\vert \,,
\end{equation}
where $A^{k}_n$ and $B_m$ are $N\times N$ matrices whose elements are
\begin{equation}
a_{ij}=\frac{-{\rm i} p^*_{j} }{p_{i}+\bar{p}_{j}} \left( -\frac{p_{i}}{p^*_{j}}\right)
^{n} \left( \frac{1- a p_{i}}{1+ a p^*_{j}}\right)^{-k} e^{\xi _{i}+\xi^*_{j}},
\quad
b_{ij}=\frac{\alpha_i \alpha^*_j }{p^*_{i}+{p}_{j}}\left( -\frac{p^*_{i}}{{p}_{j}}\right) ^{m}\,,
\end{equation}
with
\begin{equation*}
\xi_{i}= -\frac{1}{p_{i}}t +\xi _{i0},\quad \xi^*_{j}=-\frac{1}{p^*_{j}}t+\xi^*_{j0},
\end{equation*}
where $\Phi_n^{k}$, $\bar{\Phi}_n^{k}$ and $C_m$, $\bar{C}_m$ are $N$-component row vectors:
\begin{eqnarray*}
&&
\Phi^k_n =\left[p_1^n (1- a p_{1})^{-k} e^{\xi _{1}},\cdots, p_N^n  (1- a  p_{N})^{-k} e^{\xi _{N}}\right] \,,\ \
C_m=\left[\alpha_1 ({p}^*_{1})^{m},\cdots, \alpha_N ({p}^*_{N})^{m} \right],\ \
\\
&&
\bar{\Phi}^k_n=\left[(-p^*_{1})^{-n} (1+ a p^*_{1})^{k} e^{ \xi^*_{1}},\cdots ,(-p^*_{N})^{-n} (1+ a p^*_{N})^{k} e^{ \xi^*_{N}}\right],
\ \
\bar{C}_m=\left[\alpha^*_1 (-{p}_{1})^{-m},\cdots, \alpha^*_N (-{p}_{N})^{-m} \right] \,.
\end{eqnarray*}
\end{theorem}

\begin{theorem}
A fully discrete MT model is proposed as
\begin{eqnarray}
&&  \displaystyle  \frac{{\rm i}}{a}  (u^l_{k+1} - u^l_{k}) +  v^l_{k+1} +    u^l_{k+1} v^l_{k+1} \tilde{v}^l_{k}=0\,,  \label{fully-MTa} \\ [5pt]
&& \displaystyle  \frac{{\rm i}}{b}  (v^{l+1}_{k} - v^{l}_{k}) +  u^{l+1}_k +   v^{l+1}_k u^{l+1}_k \tilde{u}^{l}_k =0\,, \label{fully-MTb}
\end{eqnarray}
where $a$ and $b$ are the lattice spacing parameters, $k$ and $l$ are the discrete lattice variables along $x$- and $t$-direction, respectively,
so that $u^l_k=u(x=ka,t=lb)$, $v^l_k=v(x=ka,t=lb)$.
Fully discrete MT model (\ref{fully-MTa})-- (\ref{fully-MTb})
admits multi-soliton solution as follows
\begin{equation} \label{full-trf}
u^l_k=\frac{g^l_k}{\tilde{f}_{k}^{l}}\,, \quad
v^l_k=\frac{h^l_k}{f^l_k}, \quad
\tilde{u}^l_k=\frac{\tilde{g}^l_k}{f_{k}^{l}}\,, \quad
\tilde{v}^l_k=\frac{\tilde{h}^l_k}{\tilde{f}^l_k},
\end{equation}
where
\begin{equation}
f^l_k,=f^{0,l}_{1,k},\ \ \tilde{f}^l_k=f^{0,l}_{0,k},\ \ g^l_k= {\rm i} g^{0,l}_{-1,k} ,\ \  h^l_k=g^{0,l}_{0,k}\, ,
\ \  \tilde{g}_{k}^{l} = -{\rm i}\bar{g}_{1,k,l}^{0}, \ \ \tilde{h}_{k}^{l}=-\bar{g}_{0,k,l}^{0}.
\end{equation}
Here $f^{m,l}_{n,k}$ and $g^{m,l}_{n,k}$ are $2N \times 2N$ and $(2N+1) \times (2N+1)$ determinants defined by
\begin{equation}
f^{m,l}_{n,k} =\left\vert
\begin{array}{cc}
A^{l}_{n,k} & I \\
-I & B_m\end{array}\right\vert \,, \quad
g^{m,l}_{n,k} =
\left\vert
\begin{array}{ccc}
A^{l}_{n,k} & I & (\Phi_{n,k}^{l})^T \\
-I & B_m & \mathbf{0}^{T} \\
\mathbf{0} & -\bar{C}_m & 0\end{array}\right\vert  \,, \ \ \
\tilde{g}_{n,k}^{m,l}=\left\vert
\begin{array}{ccc}
A_{n-1,k}^l & I & \mathbf{0}^{T} \\
-I & B_{m} & {C}_{m}^{T} \\
-\bar{\Phi}^l_{n,k} & \mathbf{0} & 0\end{array}\right\vert \,,
\end{equation}
where $B_m$ and $A^{l}_{n,k}$ is a $N\times N$ matrix whose elements are
\begin{equation}
a_{ij}=\frac{-{\rm i} {p}^*_{j} }{p_{i}+{p}^*_{j}} \left( -\frac{p_{i}}{{p}^*_{j}}\right)
^{n} \left( \frac{1- a p_{i}}{1+ a {p}^*_{j}}\right)^{-k}
\left( \frac{1+ b p^{-1}_{i}}{1- b ({p}^{*}_{j})^{-1}}\right)^{-l}\,,
\quad
b_{ij}= {\frac{ 1 }{p^*_{i}+{p}_{j}}\left( -\frac{p^*_{i}}{{p}_{j}}\right) ^{m}}\,,
\end{equation}
and  $\Phi_{n,k}^{l}$ and $\bar{C}_m$ are $N$-component row vectors
\begin{eqnarray*}
&&
\Phi^l_{n,k} =\left[p_1^n (1- a p_{1})^{-k} (1+ b/ p_{1})^{-l} ,\cdots, p_N^n  (1- a  p_{N})^{-k} (1+ b /p_{N})^{-l} \right] \,, \ \
C_m=\left[ ({p}^*_{1})^{m},\cdots, ({p}^*_{N})^{m} \right],
\\
&&
\bar{\Phi}^l_{n,k} =\left[(-p^*_{1})^{-n} (1+ a p^*_{1})^{k}  (1- b/ p^*_{1})^{l},\cdots ,(-p^*_{N})^{-n} (1+ a p^*_{N})^{k} (1- b/ p^{*}_{N} )^{l} \right],\ \
\bar{C}_m=\left[ (-{p}_{1})^{-m},\cdots, (-{p}_{N})^{-m} \right] \,.
\end{eqnarray*}
\end{theorem}

 \begin{remark}
 Obviously, in the continuous limits of $b \to 0$, the fully discrete MT model (\ref{fully-MTa})-- (\ref{fully-MTb})  converges to the semi-discrete MT model (\ref{semi-MTa})-- (\ref{semi-MTb}). Moreover,  as $a \to 0$, the semi-discrete MT model (\ref{semi-MTa})-- (\ref{semi-MTb})  converge to the MT model  (\ref{MTa})-- (\ref{MTb}).
 \end{remark}

   \begin{remark}
In the discrete cases, the complex conjugate conditions are not satisfied for the bright soliton solutions. In other words, $\tilde{u}^{l}_k$, $\tilde{v}^{l}_k$ are not the complex conjugate of $u^{l}_k$,  $v^{l}_k$ and
$\tilde{u}_k(t)$, $\tilde{v}_k(t)$ are not the complex conjugate of $u_k(t)$,  $v_k(t)$. However, in the continuous limits of $a, b \to 0$ due to
\begin{eqnarray}
 && \left( \frac{1- a p_{i}}{1+ a {p}^*_{j}}\right)^{-k} \to e^{(p_i+{p}^*_{j})x },   \quad  \left( \frac{1+ b p^{-1}_{i}}{1- b ({p}^{*}_{j})^{-1}}\right)^{-l} \to e^{-(p^{-1}_i+(p^*_{j})^{-1} t }
\end{eqnarray}
by the Miwa transformation $x=ka, t=kb$, the multi-bright solutions to the discrete cases converge to the ones of the MT model  (\ref{MTa})-- (\ref{MTb}) constructed by the authors \cite{CFBilinearMT}.
 \end{remark}

\section{Semi-discrete massive Thirring model}
\subsection{Bilinearization of the semi-discrete Massive Thirring model}
We first show that the bilinearization of the semi-discrete Massive Thirring model (\ref{semi-MTa})--(\ref{semi-MTb})
\begin{eqnarray}
&& \displaystyle \frac{{\rm i}}{a}  (g_{k+1} f_{k} - g_{k} f_{k+1}) + {h_{k+1} \tilde{f}_{k}}=0 \,,  \label{MTbtBL1} \\ [5pt]
&& \displaystyle \frac{{\rm i}}{a}  (f_{k+1} \tilde{f}_{k} - f_{k} \tilde{f}_{k+1}) + {h_{k+1} \tilde{h}_{k}}=0 \,, \label{MTbtBL2}  \\ [5pt]
&&\displaystyle \mathrm{i} D_{t} h_k \cdot\tilde{f}_k   +  g_k f_k=0 \,,   \label{MTbtBL3} \\ [5pt]
&&\displaystyle  \mathrm{i} D_{t} f_k \cdot \tilde{f}_k = g_k\tilde{g}_k  \,, \label{MTbtBL4}
\end{eqnarray}
can be established by means of the dependent variable transformations (\ref{bt_tran1}).
Substituting (\ref{bt_tran1}) into (\ref{semi-MTa})--(\ref{semi-MTb}), one obtains
\begin{eqnarray}
 && \left[ \frac{{\rm i}}{a} \frac {g_{k+1} f_{k}  - g_{k} f_{k+1}}{f_k f_{k+1}} \frac{f_k}{\tilde{f}_k} +  \frac{h_{k+1}}{f_{k+1}} \right]
 +  \left[ \frac{{\rm i}}{a}   \frac {f_{k+1} \tilde{f}_{k} - f_{k} \tilde{f}_{k+1}}{\tilde{f}_k \tilde{f}_{k+1}} + \frac{h_{k+1} \tilde{h}_k}{\tilde{f}_k \tilde{f}_{k+1}}  \right] \frac{g_{k+1}}{f_{k+1}} =0,
 \end{eqnarray}
 \begin{eqnarray}
&&\left[
  \mathrm{i}  \left(\frac{h_k}{\tilde{f}_k}\right)_t \frac{\tilde{f}_k}{f_k}  +  \frac{g_k}{\tilde{f}_k}
\right]
+\left[
  \mathrm{i}  \left(\frac{\tilde{f}_k}{f_k} \right)_t
 +  \frac{g_k \tilde{g}_k}{f_k \tilde{f}_k}  \right]  \frac{h_k}{\tilde{f}_k} =0 \,,
\end{eqnarray}
{where $u_{k+1} - u_{k}=  \left( \frac{g_{k+1}}{f_{k+1}} - \frac{g_{k}}{f_{k}} \right) \frac{f_k}{\tilde{f}_k} +   \frac{g_{k+1}}{f_{k+1}} \left( \frac{f_{k+1}}{\tilde{f}_{k+1}} - \frac{f_{k}}{\tilde{f}_{k}} \right)$ . }
Bilinear equations (\ref{MTbtBL1})--(\ref{MTbtBL4}) follows by taking zero for each group inside bracket.

\subsection{Reductions from the two-component KP-Toda hierarchy}
Define the following tau-functions for two-component KP-Toda hierarchy,
\begin{equation}
f^{m,k_2}_{n,k_1} =\left\vert
\begin{array}{cc}
A^{k_1}_n & I \\
-I & B^{k_2}_m\end{array}\right\vert \,,
\end{equation}
\begin{equation}
g^{m,k_2}_{n,k_1} =
\left\vert
\begin{array}{ccc}
A^{k_1}_n & I & (\Phi_n^{k_1})^T \\
-I & B^{k_2}_m & \mathbf{0}^{T} \\
\mathbf{0} & -\bar{\Psi}^{k_2}_m & 0\end{array}\right\vert \,,\quad
\bar{g}^{m,k_2}_{n,k_1} =\left\vert
\begin{array}{ccc}
\tilde{A}^{k_1}_n & I & \mathbf{0}^{T} \\
-I & B^{k_2}_m & ({\Psi }^{k_2}_m)^{T} \\
-\bar{\Phi}^{k_1}_n & \mathbf{0} & 0\end{array}\right\vert \,,
\end{equation}
where $A^{k_1}_n$, $\tilde{A}^{k_1}_n$ and $B^{k_2}_m$ are $N\times N$ matrices whose elements are
\begin{equation}
a_{ij}=\frac{\mu \bar{p}_{j} }{p_{i}+\bar{p}_{j}} \left( -\frac{p_{i}}{\bar{p}_{j}}\right)
^{n} \left( \frac{1- a_1 p_{i}}{1+ a_1 \bar{p}_{j}}\right)^{-k_1} e^{\xi _{i}+\bar{\xi}_{j}},
\quad
\tilde{a}_{ij}=-\frac{\mu p_{i} }{p_{i}+\bar{p}_{j}}\left( -\frac{p_{i}}{\bar{p}_{j}}\right)
^{n}\left( \frac{1- a_1 p_{i}}{1+ a_1 \bar{p}_{j}}\right)^{-k_1} e^{\xi _{i}+\bar{\xi}_{j}},
\end{equation}
\begin{equation*}
b_{ij}=\frac{\nu}{q_{i}+\bar{q}_{j}}\left( -\frac{q_{i}}{\bar{q}_{j}}\right) ^{m} \left( \frac{1- a_2 q_{i}}{1+ a_2 \bar{q}_{j}}\right)^{-k_2}e^{\eta _{i}+\bar{\eta}_{j}}\,,
\end{equation*}with
\begin{equation*}
\xi _{i}=\frac{1}{p_{i}}x_{-1}+\xi _{i0},\quad \bar{\xi}_{j}=\frac{1}{\bar{p}_{j}}x_{-1}+\bar{\xi}_{j0}, \quad
\eta _{i}=\frac{1}{q_{i}}y_{-1}+\eta _{i0},\quad \bar{\eta}_{j}=\frac{1}{\bar{q}_{j}}y_{-1}+\bar{\eta}_{j0},
\end{equation*}
where $\Phi $, $\Psi $, $\bar{\Phi}$ and $\bar{\Psi}$ are $N$-component row vectors
\begin{equation*}
\Phi_n =\left[p_1^n (1- a_1 p_{1})^{-k_1} e^{\xi _{1}},\cdots, p_N^n  (1- a_1 p_{N})^{-k_1} e^{\xi _{N}}\right] \,,\
\bar{\Phi}_n=\left[(-\bar{p}_{1})^{-n} (1+ a_1 \bar{p}_{1})^{k_1} e^{\bar{\xi}_{1}},\cdots ,(-\bar{p}_{N})^{-n} (1+ a_1 \bar{p}_{N})^{k_1} e^{\bar{\xi}_{N}}\right] \,,\
\end{equation*}
\begin{equation*}
\Psi_m =\left[q_1^m (1- a_2 q_{1})^{-k_2} e^{\eta _{1}},\cdots, q_N^m  (1- a_2 q_{N})^{-k_2} e^{\eta _{N}}\right] \,,\
\bar{\Psi}_n=\left[(-\bar{q}_{1})^{-m} (1+ a_2 \bar{q}_{1})^{k_2} e^{\bar{\eta}_{1}},\cdots ,(-\bar{q}_{N})^{-m} (1+ a_2 \bar{q}_{N})^{k_2} e^{\bar{\eta}_{N}}\right] \,,\
\end{equation*}
Then the bilinear equations are satisfied
\begin{eqnarray}
\label{before-kp-1} && D_{x_{-1}}g^{m,k_2}_{n,k_1} \cdot f^{m,k_2}_{n,k_1} =  g^{m,k_2}_{n-1,k_1} f^{m,k_2}_{n+1,k_1},
\\
\label{before-kp-4} &&D_{y_{-1}} f^{m,k_2}_{n+1,k_1}  \cdot f^{m,k_2}_{n,k_1}=  -\mu \nu g^{m+1,k_2}_{n,k_1}\bar{g}^{m-1,k_2}_{n,k_1}, \\
 \label{before-kp-2}  &&\frac{g^{m+1,k_2}_{n,k_1+1}   f^{m,k_2}_{n+1,k_1} - g^{m+1,k_2}_{n,k_1}  f^{m,k_2}_{n+1,k_1+1} }{a_1}  =  {g^{m+1,k_2}_{n+1,k_1+1}  f^{m,k_2}_{n,k_1} },
\\
\label{before-kp-3}  && \frac{f^{m,k_2+1}_{n+1,k_1}  f^{m,k_2}_{n,k_1}  - f^{m,k_2}_{n+1,k_1}  f^{m,k_2+1}_{n,k_1}}{a_2}  =  { \mu \nu g^{m,k_2}_{n,k_1}    \bar{g}^{m,k_2+1}_{n,k_1}}.
\end{eqnarray}
The proof of above four bilinear equations, along with another set of four bilinear equations (\ref{fullBL1})--(\ref{fullBL4}), can be done by the determinant technique for both the continuous and discrete KP hierarchies \cite{MiyakeOhtaGram,OHTI93}, which is omitted here.
\subsection{Reductions}
In order to perform 2-reduction, we take the reduction condition
\begin{equation}\label{reduction-conditon}
q_i=\bar{p}_i,\ \ \bar{q}_i=p_i, \ \ a_1=-a_2=a.
\end{equation}
Then it can be shown that
\begin{eqnarray}
\partial_{x_{-1}} f^{m,k_2}_{n,k_1}  = \partial_{y_{-1}}  f^{m,k_2}_{n,k_1}\,, \quad
\partial_{x_{-1}} g^{m,k_2}_{n,k_1}  = \partial_{y_{-1}}  g^{m,k_2}_{n,k_1},
\end{eqnarray}
\begin{eqnarray}
f^{m+1,k_2}_{n+1,k_1}  = f^{m,k_2}_{n,k_1}\,, \quad g^{m+1,k_2}_{n+1,k_1}  = -g^{m,k_2}_{n,k_1},
\end{eqnarray}
\begin{eqnarray}
f^{m,k_2+1}_{n,k_1+1} = f^{m,k_2}_{n,k_1} , \quad g^{m,k_2+1}_{n,k_1+1} = -g^{m,k_2}_{n,k_1}.
\end{eqnarray}
Let $k_1-k_2=k$, from Eqs.(\ref{before-kp-1})--(\ref{before-kp-3}), we have the following bilinear equations
\begin{eqnarray}
\label{after-kp-1}  && D_{x_{-1}}g^{m}_{n,k} \cdot f^{m}_{n,k} =  g^{m}_{n-1,k} f^{m}_{n+1,k} = -g^{m+1}_{n,k} f^{m}_{n+1,k},
\\
\label{after-kp-4}   && D_{x_{-1}} f^{m}_{n+1,k}  \cdot f^{m}_{n,k}=  -\mu \nu g^{m+1}_{n,k}\bar{g}^{m-1}_{n,k}\,, \\
 \label{after-kp-2}  &&\frac{g^{m+1}_{n,k+1} f^{m}_{n+1,k}- g^{m+1}_{n,k} f^{m}_{n+1,k+1}}{a}  =  -g^{m}_{n,k+1}f^{m}_{n,k},
\\
\label{after-kp-3}  && \frac{f^{m}_{n+1,k} f^{m}_{n,k+1}- f^{m}_{n+1,k+1} f^{m}_{n,k}}{a}  = { -\mu \nu g^{m}_{n,k+1} \bar{g}^{m}_{n,k}},
\end{eqnarray}

Next, we consider complex conjugate by letting $\mu=-{\rm i}$, $\nu=1$, $n=m=0$  and taking
\begin{equation}
\bar{p}_i=p^*_i,\ \ \bar{\xi}_{i0}=\xi^*_{i0},\ \ e^{\eta_{i0}}=\alpha_i,\ \ e^{\bar{\eta}_{i0}}=\alpha^*_i\,.
\end{equation}
Under above conditions, we define $x_{-1}=-t$ and
\begin{equation}
f^0_{0,k}=\tilde{f}_k,\ \ f^0_{1,k}=f_k,\ \ g^0_{0,k}=h_k,\ \ \bar{g}^0_{0,k}=-\tilde{h}_k,\ \ g^1_{0,k}={\rm i}g_k,\ \ \bar{g}^{-1}_{0,k}=-{\rm i} \tilde{g}_k,
\end{equation}
Eqs. (\ref{after-kp-1})--(\ref{after-kp-3}) are reduced to Eqs.  (\ref{MTbtBL1})--(\ref{MTbtBL4}). Therefore, the reduction procedure is complete.



Example:

one-soliton solutions:
\begin{eqnarray}
&&
f_k= 1 + \frac{{\rm i}p_1 |\alpha_1|^2 }{(p_1+p^*_1)^2} \left(\frac{1-ap_1}{1+ap^*_1} \right)^{-k} e^{\xi_1+\xi^*_1}
= 1 + \frac{{\rm i}p_1 |\alpha_1|^2 }{(p_1+p^*_1)^2}  e^{\xi_1+\xi^*_1-k \ln(1-ap_1)+ k \ln(1+ap^*_1)},
\\
&&
\tilde{f}_k= 1 - \frac{{\rm i}p^*_1 |\alpha_1|^2 }{(p_1+p^*_1)^2} \left(\frac{1-ap_1}{1+ap^*_1} \right)^{-k} e^{\xi_1+\xi^*_1}
= 1 - \frac{{\rm i}p^*_1 |\alpha_1|^2 }{(p_1+p^*_1)^2}  e^{\xi_1+\xi^*_1-k \ln(1-ap_1)+ k \ln(1+ap^*_1)},
\\
&&
g_k= \frac{{\rm i}\alpha^*_1}{p_1}(1-ap_1)^{-k}e^{\xi_1}= \frac{{\rm i}\alpha^*_1}{p_1} e^{\xi_1-k \ln(1-ap_1)},\ \
\tilde{g}_k =  -\frac{{\rm i}\alpha_1}{p^*_1}(1+ap^*_1)^{k}e^{\xi^*_1} = -\frac{{\rm i}\alpha_1}{p^*_1} e^{\xi^*_1  + k \ln(1+ap^*_1)},
\\
&&
h_k= \alpha^*_1(1-ap_1)^{-k}e^{\xi_1} = \alpha^*_1 e^{\xi_1-k \ln(1-ap_1)},\ \
\tilde{h}_k = \alpha_1 (1+ap^*_1)^{k}e^{\xi^*_1}= \alpha_1 e^{\xi^*_1 + k \ln(1+ap^*_1)},
\end{eqnarray}
Define
\begin{equation}
\zeta_{i}= -\frac{1}{p_{i}}t -k \ln(1-ap_i) +\xi _{i0},\quad \tilde{\zeta}_j=-\frac{1}{p^*_{j}}t + k \ln(1+ap^*_j) +\xi^*_{j0},
\end{equation}
then
\begin{eqnarray}
&&
f_k = 1 + \frac{{\rm i}p_1 |\alpha_1|^2 }{(p_1+p^*_1)^2}  e^{\zeta_1+\tilde{\zeta}_1} ,
\\
&&
\tilde{f}_k= 1 - \frac{{\rm i}p^*_1 |\alpha_1|^2 }{(p_1+p^*_1)^2}  e^{\zeta_1+\tilde{\zeta}_1},
\\
&&
g_k= \frac{{\rm i}\alpha^*_1}{p_1} e^{\zeta_1},\ \
\tilde{g}_k  = -\frac{{\rm i}\alpha_1}{p^*_1} e^{\tilde{\zeta}_1},
\\
&&
h_k = \alpha^*_1 e^{\zeta_1},\ \
\tilde{h}_k = \alpha_1 e^{\tilde{\zeta}_1},
\end{eqnarray}


For simplicity let $\chi_1=\zeta_1+\ln\alpha^*_1$ and
\begin{equation}
  p_1= a_1+{\rm i} b_1, \ \ \chi_{10}=\xi _{10}+\ln\alpha^*_1= c_1 + {\rm i} d_1,
\end{equation}
where $a_1,b_1,c_1$ and $d_1$ are real constants. Then
\begin{eqnarray}
&&
\chi_1=\gamma_1+{\rm i}\theta_1,
\\
&&
\gamma_1=-\frac{a_1}{a^2_1+b^2_1} t - \frac{1}{2}\ln[a^2b^2_1 +(1-aa_1)^2] k + c_1, \ \
\theta_1=\frac{b_1}{a^2_1+b^2_1}t - \arctan\frac{ab_1}{1-aa_1} k+ d_1,
\end{eqnarray}
\begin{eqnarray}
&&
\tilde{\chi}_1=\tilde{\gamma}_1-{\rm i}\tilde{\theta}_1,
\\
&&
\tilde{\gamma}_1=-\frac{a_1}{a^2_1+b^2_1} t + \frac{1}{2}\ln[a^2b^2_1 +(1+aa_1)^2] k + c_1, \ \
\tilde{\theta}_1=\frac{b_1}{a^2_1+b^2_1}t - \arctan\frac{-ab_1}{1+aa_1} k + d_1,
\end{eqnarray}
then
\begin{equation}
f_k = 1 + \frac{{\rm i}(a_1+{\rm i}b_1) }{4a_1^2}  {e^{2\gamma_1}},\ \ g_k=\frac{{\rm i}}{a_1+{\rm i}b_1}e^{\gamma_1+{\rm i}\theta_1}, \ \
h_k= e^{\gamma_1+{\rm i}\theta_1}
\end{equation}

\newpage
two-soliton solutions:

\begin{eqnarray}
&&
f_k=1+c_{11^*} e^{\zeta_1+\tilde{\zeta}_1} + c_{21^*} e^{\zeta_1+\tilde{\zeta}_2}+ c_{12^*} e^{\zeta_2+\tilde{\zeta}_1}  + c_{22^*} e^{\zeta_2+\tilde{\zeta}_2}
+ c_{121^*2^*} e^{\zeta_1+\zeta_2+\tilde{\zeta}_1+\tilde{\zeta}_2},
\\
&&
\tilde{f}_k=1+c^*_{11^*} e^{\zeta_1+\tilde{\zeta}_1} + c^*_{21^*} e^{\zeta_2+\tilde{\zeta}_1}+ c^*_{12^*} e^{\zeta_1+\tilde{\zeta}_2}  + c^*_{22^*} e^{\zeta_2+\tilde{\zeta}_2}
+ c^*_{121^*2^*} e^{\zeta_1+\zeta_2+\tilde{\zeta}_1+\tilde{\zeta}_2},
\\
&&
g_k= \frac{{\rm i}\alpha^*_1}{p_1}e^{\zeta_1}+\frac{{\rm i}\alpha^*_2}{p_2}e^{\zeta_2} - \frac{{\rm i}p^*_1}{p_1p_2}c_{121^*}e^{\zeta_1+\zeta_2+\tilde{\zeta}_1}
- \frac{{\rm i}p^*_2}{p_1p_2} c_{122^*} e^{\zeta_1+\zeta_2+\tilde{\zeta}_2},
\\
&&
\tilde{g}_k= -\frac{{\rm i}\alpha_1}{p^*_1}e^{\tilde{\zeta}_1} - \frac{{\rm i}\alpha_2}{p^*_2}e^{\tilde{\zeta}_2} + \frac{{\rm i}p_1}{p^*_1p^*_2}c^*_{121^*}e^{\tilde{\zeta}_1+\tilde{\zeta}_2+\zeta_1}
+ \frac{{\rm i} p_2}{p^*_1p^*_2} c^*_{122^*} e^{\tilde{\zeta}_1+\tilde{\zeta}_2+\zeta_2},
\\
&&
h_k=\alpha^*_1e^{\zeta_1}+\alpha^*_2e^{\zeta_2} + c_{121^*}e^{\zeta_1+\zeta_2+\tilde{\zeta}_1} + c_{122^*} e^{\zeta_1+\zeta_2+\tilde{\zeta}_2},
\\
&&
\tilde{h}_k=\alpha_1e^{\tilde{\zeta}_1}+\alpha_2e^{\tilde{\zeta}_2} + c^*_{121^*}e^{\tilde{\zeta}_1+\tilde{\zeta}_2+\zeta_1} + c^*_{122^*} e^{\tilde{\zeta}_1+\tilde{\zeta}_2+ \zeta_2},
\end{eqnarray}

with
\begin{eqnarray}
&&
c_{ij^*} =  \frac{{\rm i}\alpha^*_i\alpha_j p_i}{(p_i+p^*_j)},\ \
\\
&&
c_{121^*2^*} = |p_1-p_2|^2\left[ \frac{c_{11^*}c_{22^*}}{(p_1+p^*_2)(p^*_1+p_2)} - \frac{c_{12^*}c_{21^*}}{(p_1+p^*_1)(p_2+p^*_2)}\right]
\\
&&
c_{12i^*} = (p_2-p_1)p^*_i\left[ \frac{\alpha^*_2c_{1i^*}}{p_1(p_2+p^*_i)} - \frac{\alpha^*_1 c_{2i^*}}{p_2(p_1+p^*_i)} \right]
\end{eqnarray}

\begin{figure}[!htbp]
\centering
\subfigure[]{\includegraphics[height=2.0in,width=2.7in]{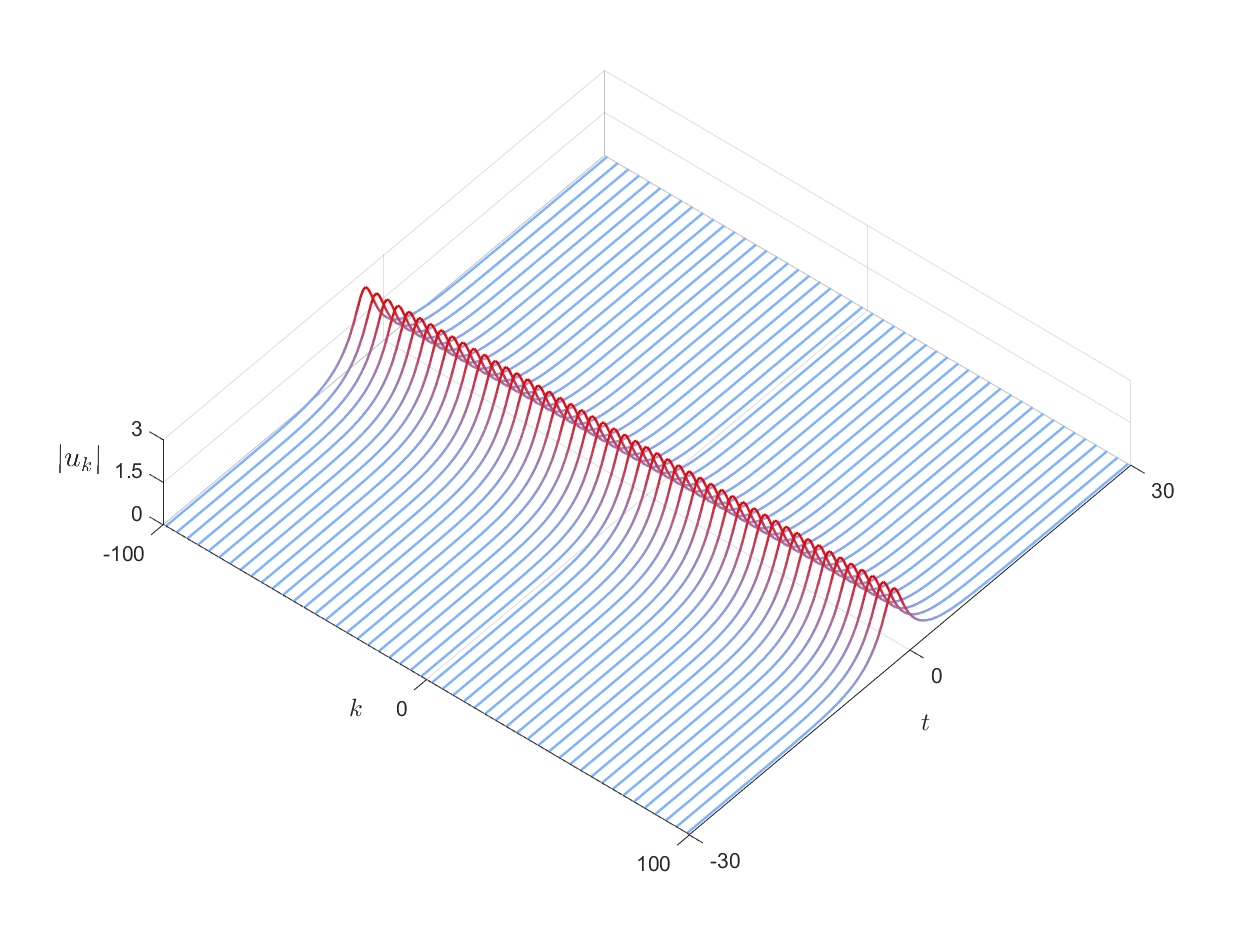}}\hspace{0.1cm}
\subfigure[]{\includegraphics[height=2.0in,width=2.7in]{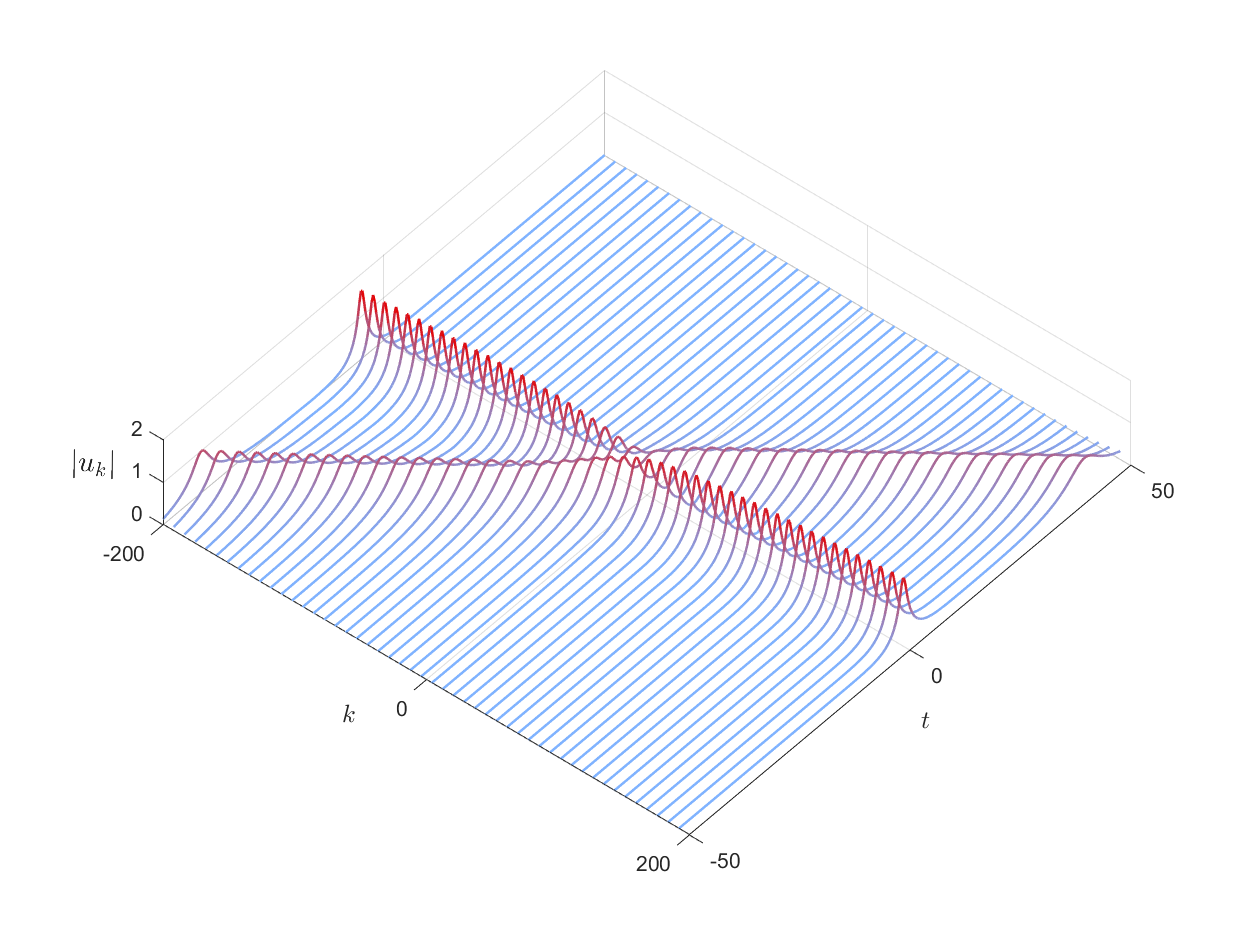}}
\caption{The soltion solutions of the semi-discrete Massive Thirring model: (a) one-soliton with $p_1=2+{\rm i}$, $\alpha_1=1$, $a=10^{-3}$ and $\xi_{1,0}=0$; (b) two-soliton with $p_1=1+{\rm i}$, $p_2=4+2{\rm i}$, $\alpha_1=1$, $\alpha_2=3+7{\rm i}$, $a=10^{-2}$ and $\xi_{1,0}=\xi_{2,0}=0$.
\label{fig-semi}}
\end{figure}

\section{Fully discrete massive Thirring model}
\subsection{Bilinearization of the fully discrete MT model}
By virtue of  variable transformations (\ref{full-trf}),
it is easily shown
\begin{eqnarray*}
   && u^l_{k+1}-u^l_k
   =\frac{g^l_{k+1}}{\tilde{f}_{k+1}^{l}} - \frac{g^l_k}{\tilde{f}_{k}^{l}}
    =\frac{g^l_{k+1}f^l_k- g^l_{k}f^l_{k+1}} {f^l_{k+1} f_{k}^{l}} \frac{f_{k}^{l}}{\tilde{f}_{k}^{l}} +\frac{g^l_{k+1}}{\tilde{f}_{k+1}^{l}}
   \left(
   \frac{f^l_{k+1} \tilde{f}^l_{k} - f^l_{k} \tilde{f}^l_{k+1}}
   {f^l_{k+1} \tilde{f}^l_{k}}
   \right)\,.
   \end{eqnarray*}
Substituting into (\ref{fully-MTa}), one obtains
 \begin{eqnarray}
 && \left[ \frac{{\rm i}}{a} \frac {g^l_{k+1} f^l_{k}  - g^l_{k} f^l_{k+1}}{f^l_k f^l_{k+1}} \frac{f^l_k}{\tilde{f}^l_k} +  \frac{h^l_{k+1}}{f^l_{k+1}} \right]
 +  \left[ \frac{{\rm i}}{a}   \frac {f^l_{k+1} \tilde{f}^l_{k} - f^l_{k} \tilde{f}^l_{k+1}}{\tilde{f}^l_k \tilde{f}^l_{k+1}} + \frac{h^l_{k+1} \tilde{h}^l_k}{\tilde{f}^l_k \tilde{f}^l_{k+1}}  \right] \frac{g^l_{k+1}}{f^l_{k+1}} =0.
 \end{eqnarray}


By taking zero in each group, we have
\begin{eqnarray}
\label{fullBL1a} && \frac{\rm i} {a} (g^l_{k+1} f^l_{k} - g^l_{k} f^l_{k+1})+ h^l_{k+1} {\tilde{f}}^l_{k} =0\,,
\\
\label{fullBL2a}  &&  \frac{\rm i} {a} \left[f^l_{k+1} \tilde{f}^l_{k} - f_{k} {\tilde{f}}^l_{k+1} \right] +h^l_{k+1} {\tilde{h}}^l_{k}=0.
\end{eqnarray}
Similarly, we can obtain
\begin{eqnarray}
\label{fullBL3a}  && \frac{\rm i} {b} \left[h_{k}^{l+1}   {\tilde{f}}_{k}^{l} - h_{k}^{l}  {\tilde{f}}_{k}^{l+1} \right] +  g_{k}^{l+1}  f_{k}^{l}=0,
\\
\label{fullBL4a}  && \frac{\rm i} {b} \left[f_{k}^{l+1}  {\tilde{f}}_{k}^{l} - f_{k}^{l}  {\tilde{f}}_{k}^{l+1}\right] =   g_{k}^{l+1}  {\tilde{g}}_{k}^{l} \,.
\end{eqnarray}
Eqs. (\ref{fullBL1a})--(\ref{fullBL4a}) constitutes the bilinear equations of the fully discrete MT model.
\subsection{Reductions from  the two-component discrete KP-Toda hierarchy}
In what follows, we construct multi-soliton solution to the fully discrete MT model via the KP reduction method.
To this end, define the following tau-functions of two-component discrete KP-Toda hierarchy,
\begin{equation}
f_{n,k_1,l_1}^{m,k_2,l_2}=\left\vert
\begin{array}{cc}
A_{n,k_1,l_1}& I \\
-I & B_{m,k_2,l_2}  \end{array}\right\vert \,,
\end{equation}
\begin{equation}
g_{n,k_1,l_1}^{m,k_2,l_2}=
\left\vert
\begin{array}{ccc}
A_{n,k_1,l_1}& I & \Phi^T_{n,k_1,l_1}  \\
-I & B_{m,k_2,l_2}  & \mathbf{0}^{T} \\
\mathbf{0} & -\bar{\Psi}_{m,k_2,l_2} & 0\end{array}\right\vert \,,\quad
\bar{g}_{n,k_1,l_1}^{m,k_2,l_2}=\left\vert
\begin{array}{ccc}
\tilde{A}_{n,k_1,l_1} & I & \mathbf{0}^{T} \\
-I & B_{m,k_2,l_2} & {\Psi }_{m,k_2,l_2}^{T} \\
-\bar{\Phi}_{n,k_1,l_1} & \mathbf{0} & 0\end{array}\right\vert \,,
\end{equation}
where $A_{n,k_1,l_1}$ ,$\tilde{A}_{n,k_1,l_1}$ and $B_{m,k_2,l_2}$ are $N\times N$ matrices whose elements are
\begin{equation*}
a_{ij}=\frac{\mu \bar{p}_{j} }{p_{i}+\bar{p}_{j}}\left( -\frac{p_{i}}{\bar{p}_{j}}\right)
^{n} \left( \frac{1- a_1 p_{i}}{1+ a_1 \bar{p}_{j}}\right)^{-k_1}  \left( \frac{1- b_1 p^{-1}_{i}}{1+ b_1 \bar{p}^{-1}_{j}}\right)^{-l_1},
\end{equation*}
\begin{equation*}
\tilde{a}_{ij}=-\frac{\mu p_{i} }{p_{i}+\bar{p}_{j}}\left( -\frac{p_{i}}{\bar{p}_{j}}\right)
^{n} \left( \frac{1- a_1 p_{i}}{1+ a_1 \bar{p}_{j}}\right)^{-k_1}  \left( \frac{1- b_1 p^{-1}_{i}}{1+ b_1 \bar{p}^{-1}_{j}}\right)^{-l_1}
\end{equation*}
\begin{equation*}
b_{ij}=\frac{\nu}{q_{i}+\bar{q}_{j}}\left( -\frac{q_{i}}{\bar{q}_{j}}\right) ^{m}\left( \frac{1- a_2 q_{i}}{1+ a_2 \bar{q}_{j}}\right)^{-k_2} \left( \frac{1- b_2 q^{-1}_{i}}{1+ b_2 \bar{q}^{-1}_{j}}\right)^{-l_2}
\,,
\end{equation*}
and $\Phi $, $\Psi $, $\bar{\Phi}$ and $\bar{\Psi}$ are $N$-component row vectors
\begin{equation*}
\Phi_{n,k_1,l_1} =\left[p_1^n (1- a_1 p_{1})^{-k_1}  (1- b_1 p^{-1}_{1})^{-l_1},\cdots ,p_N^n (1- a_1 p_{N})^{-k_1}  (1- b_1 p^{-1}_{1})^{-l_1} \right] \,,
\end{equation*}
\begin{equation*}
\bar{\Phi}_{n,k_1,l_1} =\left[(-\bar{p}_{1})^{-n} (1+ a_1 \bar{p}_{1})^{k_1}  (1+ b_1 \bar{p}^{-1}_{1})^{l_1},\cdots ,(-\bar{p}_{N})^{-n} (1+ a_1 \bar{p}_{N})^{k_1}
(1+ b_1 \bar{p}^{-1}_{N})^{l_1} \right] \,,\
\end{equation*}
\begin{equation*}
\Psi_{m,k_2,l_2} =\left[q_1^m (1- a_2 q_{1})^{-k_2}  (1- b_2 q^{-1}_{1})^{-l_2},\cdots,q_N^m (1- a_2 q_{N})^{-k_2}  (1- b_2 q^{-1}_{1})^{-l_2} \right] \,,
\end{equation*}
\begin{equation*}
\bar{\Psi}_{m,k_2,l_2} =\left[(-\bar{q}_{1})^{-m} (1+ a_2 \bar{q}_{1})^{k_2}  (1+ b_2 \bar{q}^{-1}_{1})^{l_2},\cdots ,(-\bar{q}_{N})^{-m} (1+ a_2 \bar{q}_{N})^{k_2}
(1+ b_2 \bar{q}^{-1}_{N})^{l_2} \right] \,.
\end{equation*}
Then it can shown that the following discrete bilinear equations are satisfied
\begin{eqnarray}
 \label{fullBL1} && \frac{1}{a_1} \left(g_{n,k_1+1,l_1}^{m+1,k_2,l_2}  f_{n+1,k_1,l_1}^{m,k_2,l_2}- g_{n,k_1,l_1}^{m+1,k_2,l_2}  f_{n+1,k_1+1,l_1}^{m,k_2,l_2}\right)  = g_{n+1,k_1+1,l_1}^{m+1,k_2,l_2} f_{n,k_1,l_1}^{m,k_2,l_2},
\\
 \label{fullBL2} &&  \frac{1}{a_2} \left( f^{m,k_2+1,l_2}_{n+1,k_1,l_1} f^{m,k_2,l_2}_{n,k_1,l_1}- f^{m,k_2,l_2}_{n+1,k_1,l_1} f^{m,k_2+1,l_2}_{n,k_1,l_1}\right)  =  \mu \nu  g_{n,k_1,l_1}^{m,k_2,l_2}  \bar{g}_{n,k_1,l_1}^{m,k_2+1,l_2},\\
  \label{fullBL3} &&\frac{1}{b_1} \left(g_{n,k_1,l_1+1}^{m,k_2,l_2} f_{n,k_1,l_1}^{m,k_2,l_2}- g_{n,k_1,l_1}^{m,k_2,l_2} f_{n,k_1,l_1+1}^{m,k_2,l_2}\right)  =  g_{n-1,k_1,l_1+1}^{m,k_2,l_2}f_{n+1,k_1,l_1}^{m,k_2,l_2},
\\
 \label{fullBL4} &&\frac{1}{b_2} \left(f_{n+1,k_1,l_1}^{m,k_2,l_2+1} f_{n,k_1,l_1}^{m,k_2,l_2}- f_{n+1,k_1,l_1}^{m,k_2,l_2} f_{n,k_1,l_1}^{m,k_2,l_2+1}\right)  =  -\mu \nu  g_{n,k_1,l_1}^{m+1,k_2,l_2}
\bar{g}_{n,k_1,l_1}^{m-1,k_2,l_2+1}\,.
\end{eqnarray}
\subsection{Reductions}
To perform the period 2 reduction, we impose the reduction conditions
\begin{equation}
q_i=\bar{p}_i,\ \ \bar{q}_i=p_i, \ \ a_1=-a_2=a,\ \ {b_1=-b_2=-b}.
\end{equation}
Then we can show that
\begin{eqnarray}
f^{m+1,k_2,l_2}_{n+1,k_1,l_1}  = f^{m,k_2,l_2}_{n,k_1,l_1}\,, \quad g^{m+1,k_2,l_2}_{n+1,k_1,l_1}  = -g^{m,k_2,l_2}_{n,k_1,l_1}, \ \
\bar{g}^{m+1,k_2,l_2}_{n+1,k_1,l_1}  = -\bar{g}^{m,k_2,l_2}_{n,k_1,l_1}
\end{eqnarray}
\begin{eqnarray}
f^{m,k_2+1,l_2}_{n,k_1+1,l_1} = f^{m,k_2,l_2}_{n,k_1,l_1} , \quad g^{m,k_2+1,l_2}_{n,k_1+1,l_1} =  g^{m,k_2,l_2}_{n,k_1,l_1},  \quad \bar{g}^{m,k_2+1,l_2}_{n,k_1+1,l_1} =  \bar{g}^{m,k_2,l_2}_{n,k_1,l_1},
\end{eqnarray}
and
\begin{eqnarray}
f_{n,k_1,l_1+1}^{m,k_2,l_2+1} = f_{n,k_1,l_1}^{m,k_2,l_2}, \quad g_{n,k_1,l_1+1}^{m,k_2,l_2+1} = g_{n,k_1,l_1}^{m,k_2,l_2},\quad \bar{g}_{n,k_1,l_1+1}^{m,k_2,l_2+1} = \bar{g}_{n,k_1,l_1}^{m,k_2,l_2}.
\end{eqnarray}

Then it can shown that the following discrete bilinear equations are satisfied
\begin{eqnarray}
 \label{fullBL11} && \frac{1}{a_1} \left(g_{n,k_1+1,l_1}^{m+1,k_2,l_2}  f_{n+1,k_1,l_1}^{m,k_2,l_2}- g_{n,k_1,l_1}^{m+1,k_2,l_2}  f_{n+1,k_1+1,l_1}^{m,k_2,l_2}\right)  = g_{n+1,k_1+1,l_1}^{m+1,k_2,l_2} f_{n,k_1,l_1}^{m,k_2,l_2},
\\
 \label{fullBL12} &&  \frac{1}{a_2} \left( f^{m,k_2+1,l_2}_{n+1,k_1,l_1} f^{m,k_2,l_2}_{n,k_1,l_1}- f^{m,k_2,l_2}_{n+1,k_1,l_1} f^{m,k_2+1,l_2}_{n,k_1,l_1}\right)  =  \mu \nu  g_{n,k_1,l_1}^{m,k_2,l_2}  \bar{g}_{n,k_1,l_1}^{m,k_2+1,l_2},\\
  \label{fullBL13} &&\frac{1}{b_1} \left(g_{n,k_1,l_1+1}^{m,k_2,l_2} f_{n,k_1,l_1}^{m,k_2,l_2}- g_{n,k_1,l_1}^{m,k_2,l_2} f_{n,k_1,l_1+1}^{m,k_2,l_2}\right)  =  g_{n-1,k_1,l_1+1}^{m,k_2,l_2}f_{n+1,k_1,l_1}^{m,k_2,l_2},
\\
 \label{fullBL14} &&\frac{1}{b_2} \left(f_{n+1,k_1,l_1}^{m,k_2,l_2+1} f_{n,k_1,l_1}^{m,k_2,l_2}- f_{n+1,k_1,l_1}^{m,k_2,l_2} f_{n,k_1,l_1}^{m,k_2,l_2+1}\right)  =  -\mu \nu  g_{n,k_1,l_1}^{m+1,k_2,l_2}
\bar{g}_{n,k_1,l_1}^{m-1,k_2,l_2+1}\,.
\end{eqnarray}
By utilizing these relations, and taking $k_1-k_2=k$, $l_1-l_2=l$,
Eqs. (\ref{fullBL11})--(\ref{fullBL14}) lead to
\begin{eqnarray}
&&\label{fullBL1ab} \frac{1}{a} \left(- g_{n-1,k+1,l}^{m}  f_{n+1,k,l}^{m} + g_{n-1,k,l}^{m}  f_{n+1,k+1,l}^{m}\right)  = - g_{n,k+1,l}^{m} f_{n,k,l}^{m},
\\
&& \label{fullBL2ab}  -\frac{1}{a} \left( f^{m}_{n+1,k-1,l} f^{m}_{n,k,l}- f^{m}_{n+1,k,l} f^{m}_{n,k-1,l}\right)  =  \mu \nu  g_{n,k,l}^{m}  \bar{g}_{n,k-1,l}^{m},\\
&& \label{fullBL3ab} -\frac{1}{b} \left(g_{n,k,l+1}^{m} f_{n,k,l}^{m}- g_{n,k,l}^{m} f_{n,k,l+1}^{m}\right)  =  g_{n-1,k,l+1}^{m }f_{n+1,k,l}^{m},
\\
&& \label{fullBL4ab} \frac{1}{b} \left(f_{n+1,k,l-1}^{m} f_{n,k,l}^{m}- f_{n+1,k,l}^{m} f_{n,k,l-1}^{m}\right)  =  -\mu \nu  g_{n-1,k,l}^{m}
\bar{g}_{n+1,k,l-1}^{m}\,.
\end{eqnarray}

Next, we consider complex conjugate by letting $\mu=-{\rm i}$, $\nu=1$ and  $\bar{p}_i=p^*_i$. Further, we define
\begin{equation}
f_{1,k,l}^{0} = f_{k}^{l}, \ \ f_{0,k,l}^{0}=\tilde{f}_{k}^{l},
\ \ { {g}_{-1,k,l}^{0}=- {\rm i}g_{k}^{l}, \ \ \bar{g}_{1,k,l}^{0}= {\rm i} \tilde{g}_{k}^{l}, \ \
{g}_{0,k,l}^{0}=h_{k}^{l}, \ \ \bar{g}_{0,k,l}^{0}=  -\tilde{h}_{k}^{l}}
\end{equation}
Then, we have exactly the bilinear equations for the fully discrete MT model (\ref{fullBL1a})--(\ref{fullBL4a}) by letting $n=m=0$ in (\ref{fullBL1ab})--(\ref{fullBL4ab})

{ Examples for one- and two-soliton solutions of the fully discrete MT model are plotted in Figs. 1 and 2, respectively.}

\newpage

one-soliton:

\begin{eqnarray}
&&
f_{k}^{l}= 1+ \frac{{\rm i}p_1}{(p_1+p^*_1)^2}\left( \frac{1-ap_1}{1+ap^*_1} \right)^{-k}\left( \frac{1+\frac{b}{p_1}}{1-\frac{b}{p^*_1}} \right)^{-l}= 1+ \frac{{\rm i}p_1}{(p_1+p^*_1)^2} e^{\xi_1+\tilde{\xi}_1},
\\
&&
\tilde{f}_{k}^{l}= 1 - \frac{{\rm i}p^*_1}{(p_1+p^*_1)^2}\left( \frac{1-ap_1}{1+ap^*_1} \right)^{-k}\left( \frac{1+\frac{b}{p_1}}{1-\frac{b}{p^*_1}} \right)^{-l} = 1 - \frac{{\rm i}p^*_1}{(p_1+p^*_1)^2} e^{\xi_1+\tilde{\xi}_1},
\\
&&
g_{k}^{l}= \frac{{\rm i}}{p_1} \left(1-ap_1 \right)^{-k} \left( 1+\frac{b}{p_1} \right)^{-l}= \frac{{\rm i}}{p_1} e^{\xi_1},
\\
&&
\tilde{g}_{k}^{l}= -\frac{{\rm i}}{p^*_1} \left( 1+ap^*_1 \right)^{k} \left( 1-\frac{b}{p^*_1} \right)^{l} = -\frac{{\rm i}}{p^*_1} e^{\tilde{\xi}_1},
\\
&&
h_{k}^{l}=   \left(1-ap_1 \right)^{-k} \left( 1+\frac{b}{p_1} \right)^{-l} =e^{\xi_1},
\\
&&
\tilde{h}_{k}^{l}=   \left( 1+ap^*_1 \right)^{k} \left( 1-\frac{b}{p^*_1} \right)^{l} =e^{\tilde{\xi}_1},
\end{eqnarray}
with
\begin{equation*}
\xi_i= -k \ln(1-ap_i) - l\ln(1+ \frac{b}{p_i}),\ \ \tilde{\xi}_i= k \ln(1+ap^*_i) + l\ln(1- \frac{b}{p^*_i})
\end{equation*}

two-soliton solutions:

\begin{eqnarray}
&&
f_k=1+c_{11^*} e^{\zeta_1+\tilde{\zeta}_1} + c_{21^*} e^{\zeta_1+\tilde{\zeta}_2}+ c_{12^*} e^{\zeta_2+\tilde{\zeta}_1}  + c_{22^*} e^{\zeta_2+\tilde{\zeta}_2}
+ c_{121^*2^*} e^{\zeta_1+\zeta_2+\tilde{\zeta}_1+\tilde{\zeta}_2},
\\
&&
\tilde{f}_k=1+c^*_{11^*} e^{\zeta_1+\tilde{\zeta}_1} + c^*_{21^*} e^{\zeta_2+\tilde{\zeta}_1}+ c^*_{12^*} e^{\zeta_1+\tilde{\zeta}_2}  + c^*_{22^*} e^{\zeta_2+\tilde{\zeta}_2}
+ c^*_{121^*2^*} e^{\zeta_1+\zeta_2+\tilde{\zeta}_1+\tilde{\zeta}_2},
\\
&&
g_k= \frac{{\rm i} }{p_1}e^{\zeta_1}+\frac{{\rm i} }{p_2}e^{\zeta_2} - \frac{{\rm i}p^*_1}{p_1p_2}c_{121^*}e^{\zeta_1+\zeta_2+\tilde{\zeta}_1}
- \frac{{\rm i}p^*_2}{p_1p_2} c_{122^*} e^{\zeta_1+\zeta_2+\tilde{\zeta}_2},
\\
&&
\tilde{g}_k= -\frac{{\rm i} }{p^*_1}e^{\tilde{\zeta}_1} - \frac{{\rm i} }{p^*_2}e^{\tilde{\zeta}_2} + \frac{{\rm i}p_1}{p^*_1p^*_2}c^*_{121^*}e^{\tilde{\zeta}_1+\tilde{\zeta}_2+\zeta_1}
+ \frac{{\rm i} p_2}{p^*_1p^*_2} c^*_{122^*} e^{\tilde{\zeta}_1+\tilde{\zeta}_2+\zeta_2},
\\
&&
h_k= e^{\zeta_1}+ e^{\zeta_2} + c_{121^*}e^{\zeta_1+\zeta_2+\tilde{\zeta}_1} + c_{122^*} e^{\zeta_1+\zeta_2+\tilde{\zeta}_2},
\\
&&
\tilde{h}_k= e^{\tilde{\zeta}_1}+ e^{\tilde{\zeta}_2} + c^*_{121^*}e^{\tilde{\zeta}_1+\tilde{\zeta}_2+\zeta_1} + c^*_{122^*} e^{\tilde{\zeta}_1+\tilde{\zeta}_2+ \zeta_2},
\end{eqnarray}

with
\begin{eqnarray}
&&
c_{ij^*} =  \frac{{\rm i}  p_i}{(p_i+p^*_j)},\ \
\\
&&
c_{121^*2^*} = |p_1-p_2|^2\left[ \frac{c_{11^*}c_{22^*}}{(p_1+p^*_2)(p^*_1+p_2)} - \frac{c_{12^*}c_{21^*}}{(p_1+p^*_1)(p_2+p^*_2)}\right]
\\
&&
c_{12i^*} = (p_2-p_1)p^*_i\left[ \frac{ c_{1i^*}}{p_1(p_2+p^*_i)} - \frac{  c_{2i^*}}{p_2(p_1+p^*_i)} \right]
\end{eqnarray}

\begin{figure}[!htbp]
\centering
\subfigure[]{\includegraphics[height=2.0in,width=2.7in]{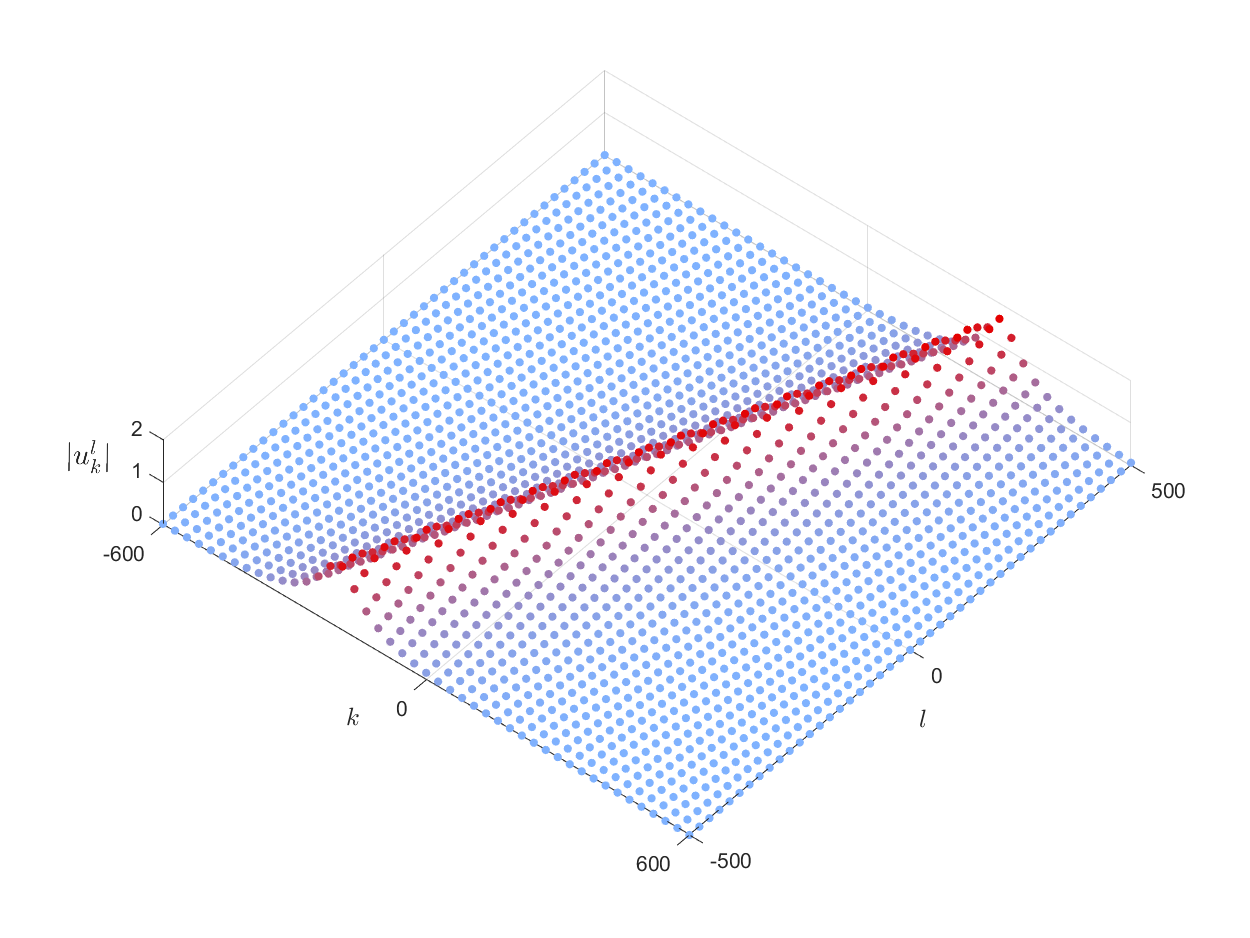}}\hspace{0.1cm}
\subfigure[]{\includegraphics[height=2.0in,width=2.7in]{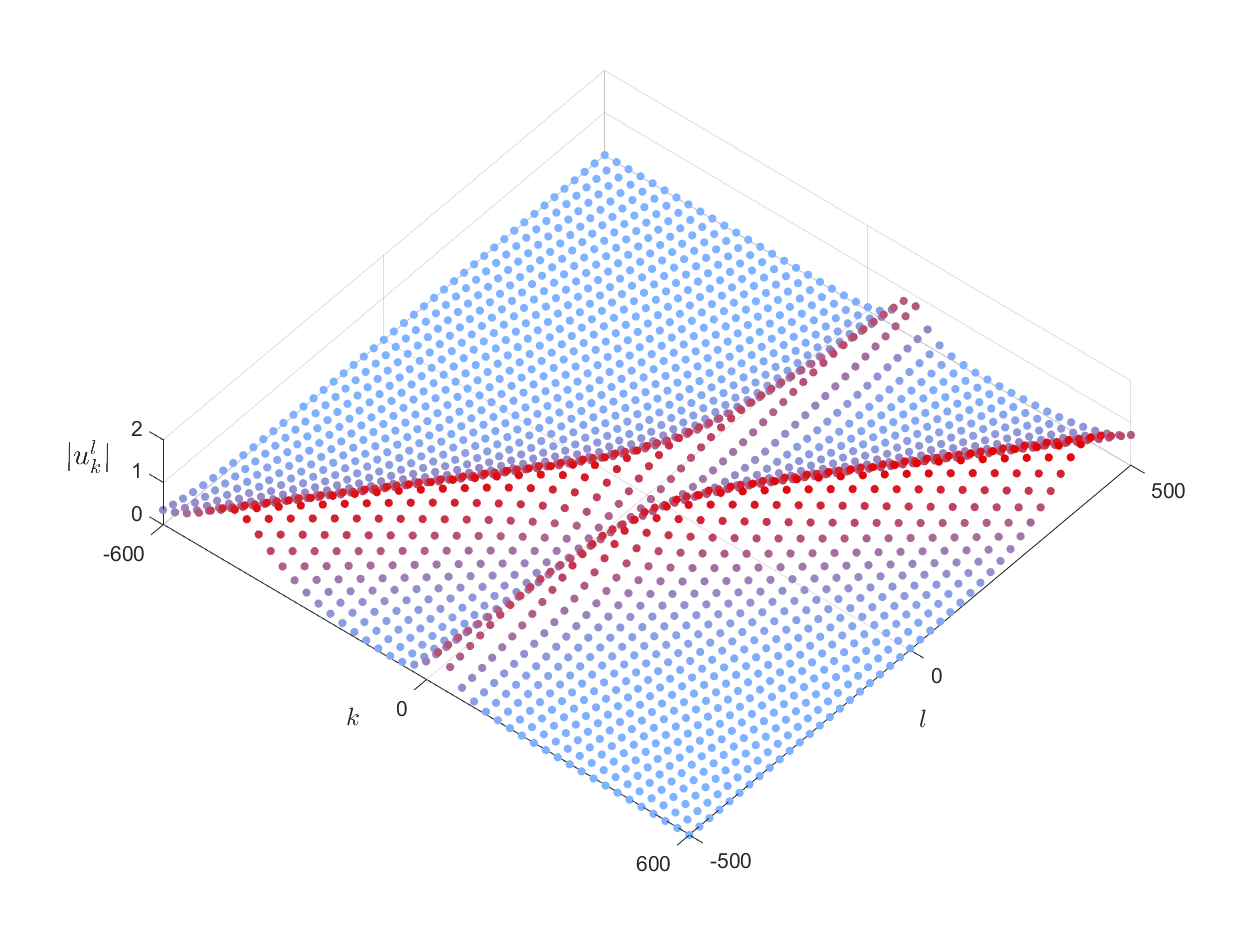}}
\caption{The soltion solutions of the full-discrete Massive Thirring model: (a) one-soliton with $p_1=1+{\rm i}$ and $a=b=10^{-2}$; (b) two-soliton with $p_1=1$, $p_2=3-\frac{1}{20}{\rm i}$ and $a=b=10^{-2}$.
\label{fig-full}}
\end{figure}

\newpage


\section{Concluding Remarks}
Starting from the two-component KP-Toda hierarchy and its discrete counterpart, we construct semi- and fully discrete analogies of classical massive Thirring model in light cone coordinates.
The integrability is guaranteed by the bilinear equations along with multi-soliton solution.

Several open problems and questions arise naturally. First, how about the symmetries including the master symmetry and the B\"acklund transformation of the discrete models? Second, could we construct the same discrete models under NVBC based on the discrete KP equation, as well as its dark, breather and rogue wave solutions?  We expect our results can draw some attentions in the study of discrete MT models.
\section*{Acknowledgement}
We thank Prof. Dmitry Pelinovsky for drawing us attention of the massive Thirring model.
JC's work was supported from the National Natural
Science Foundation of China (NSFC) (No. 12375003).
BF's work is partially supported by  the U.S. Department of Defense (DoD), Air Force for Scientific
Research (AFOSR) under grant No. W911NF2010276.


\end{document}